\documentclass[12pt,preprint]{aastex}
\usepackage{hyperref}
\def\apjs{{\rm ApJS}}                  
\def\apjl{{\rm ApJ Let}}               

\def\a4{\hsize 17.0cm \vsize 25.cm}
\newcommand{\der}[2]  { \frac{{\rm d}#1}{{\rm d}#2} }

\shorttitle{}

\begin{document}

\title{On the origin of the absorption and emission line components in the spectra of {\it PHL~293B}}

\author{Guillermo Tenorio-Tagle\altaffilmark{1}, 
Sergiy Silich\altaffilmark{1}, 
Sergio Mart\'inez-Gonz\'alez\altaffilmark{1},  
Roberto Terlevich\altaffilmark{1,2} and Elena 
Terlevich\altaffilmark{1}}

\altaffiltext{1}{Instituto Nacional de Astrof\'\i sica \'Optica y
Electr\'onica, AP 51, 72000 Puebla, M\'exico; gtt@inaoep.mx}
\altaffiltext{2}{nstitute of Astronomy, University of Cambridge, Madingley
Road, Cambridge CB3 0HA, UK}

\begin{abstract}
From the structure of {\it PHL~293B} and the physical properties  of its  
ionizing cluster and based on results of hydrodynamic models, we point at  
the various events required to explain in detail the emission and absorption 
components seen in its optical spectrum.  We ascribe the narrow and well 
centered emission lines, showing the low metallicity of the galaxy, to an HII 
region that spans through the main body of the galaxy.  The broad emission line components are due to two off-centered 
supernova remnants evolving within the ionizing cluster volume and the absorption line profiles are due to a 
stationary cluster wind able to recombine at a close distance from the cluster surface, as originally suggested by 
\citet{Silichetal2004}. Our numerical models and analytical estimates confirm the ionized and neutral column 
density values and the inferred X-ray emission derived from the observations. 
  \end{abstract}


\section{Introduction}
\label{sec:1}

{\it PHL~293B} is an extreme emission line blue compact dwarf or HII galaxy 
first listed in the ``Palomar-Haro-Luyten'' survey of faint blue  objects.
It ranks among the lowest luminosity and metallicity and smallest size 
galaxies. The metal abundances are  less than one tenth of the solar value 
\citep{French1980,Izotovetal2007,Asplundetal2009} which situates it in the 
borderline of what is considered to be an extremely metal deficient galaxy 
\citep[e.g.][]{KunthOstlin2000}. It is very compact with an effective radius 
of 0.4 kpc \citep{Cairosetal2001,Gehaetal2006} and its absolute magnitude 
according to the Sloan Digital Sky Survey (SDSS) is M$_g$ = -14.8.

The optical spectrum of {\it PHL~293B} is that of a typical HII galaxy with 
strong narrow emission lines resembling an HII region. In addition the 
spectrum shows some intriguing features: low intensity broad wings in the 
Balmer series plus weak narrow absorptions in the hydrogen recombination
lines and FeII multiplet 42, all of them blue-shifted by about 800~km s$^{-1}$
with respect to the galaxy reference frame provided by the stellar IR CaII 
triplet.

While the observed broad wings in the Balmer lines can be originated in a very
old SN  remnant (the supernova rate of {\it PHL~293B} is about $3 \times 
10^{-4}$~yr$^{-1}$),  in order to generate relatively narrow absorption 
profiles such as the ones detected in {\it PHL~293B}, it was argued that the 
absorbing material must cover a substantial fraction of the continuum source, 
in this case a young stellar cluster several parsecs in size 
\citep{Terleetal2014}.

From the observed Balmer line luminosity and using the equivalent width of the
Balmer emission lines as an age indicator, the resulting upper limit for the
ionizing star cluster mass is $M_{SC} \approx (2-4) \times 10^5$ M$_{\sun}$ 
and the upper limit for the age is about $(5-7) \times 10^6$ yr.

\citet{Terleetal2014} analyzed extensive published  and own photometric and 
spectroscopic data and found that at 3 $\sigma$ level any long term 
variability, i.e. over a few years, cannot be larger than 0.02 magnitudes and
 that any medium term variability, i.e. over a few months, should be less than
0.04 magnitudes. Furthermore there is no variability at the level of a few
tenths of magnitude over a period of 25 years. This lack of variability
suggests that the spectrum of {\it PHL~293B} is not due to transient
phenomena like a luminous blue variable  or a type IIn SN.

From the ACIS-Chandra X-ray image of {\it PHL~293B}, \citet{Terleetal2014}
estimated a luminosity upper limit in the 0.4 to 6 keV energy range of about 
$2\times 10^{38}$ erg\,s$^{-1}$. Analysis of the detected narrow absorption 
lines of Fe allows to estimate that the column densities of ionized hydrogen 
($N(HII)$) are around $10^{20}$~cm$^{-2}$ if Fe/H is similar to that of the 
ISM in {\it PHL~293B} or substantially lower if Fe/H is that of the cluster 
wind. Assuming that the cluster wind has solar abundance, the estimated 
($N(HII)$) column density is $10^{19}$~cm$^{-2}$. From the equivalent 
width of the Balmer lines \citet{Terleetal2014} also estimated that the column 
density of neutral hydrogen is $10^{13}$~cm$^{-2} <$ N(HI) 
$< 10^{14}$~cm$^{-2}$.  The true column density of neutral hydrogen however
may be higher because only Balmer absorption were used in this estimate. 
Measurements on HST images indicate that the diameter of the brightest 
cluster is less than 5 pc.

As discussed by \citet{Terleetal2014},
while the presence of broad Balmer lines and 
blue-shifted absorptions may be explained by either an extremely luminous blue 
variable (LBV) or an old SN type IIn, the lack of variability over at least a 
quarter of a century strongly argues against these scenarios. Thus the 
understanding of this object constitutes a challenge. Here we present a new 
scenario supported by detailed hydrodynamic calculations of a strongly radiative stationary
cluster wind  and a comparison of the theoretical predictions with 
the observed parameters.

Based on the structure of {\it PHL~293B} and on the physical properties of 
its ionizing cluster, we point at the various events required to explain in 
detail the emission and absorption components seen in its spectrum. In section
\ref{sec:2} we ascribe the narrow and  centered emission lines, to an HII region that 
spans through the main body of the galaxy and propose an explanation for the 
observed broad emission line components and the P-Cygni absorption line 
profiles. Section \ref{sec:3} deals with our conclusions and further predictions from 
the model.
 
\section{Negative star formation feedback in {\it PHL~293B}}
\label{sec:2}

Knowledge of the cluster mass and age allow one to estimate its mechanical
 luminosity, and the production of UV photons. These are well known negative 
star formation feedback agents that structure the galaxy ISM in a wide variety
of ways.  These parameters together with the cluster size have been shown to 
define whether the mass reinserted by massive stars ends up, after 
thermalization, streaming all supersonically as a stationary cluster wind or 
if instead, strong thermal instabilities  may deplete the pressure of the 
reinserted matter, particularly within the densest central volume, inhibiting 
its exit as part of the cluster wind. In the latter case,  the reinserted 
matter is forced to accumulate there until its mass surpasses the Jeans limit,
leading to a new stellar generation.   As shown before, the more massive 
a young cluster is, the more of the reinserted matter  is to  accumulate 
and the smallest the resultant fraction of its mechanical energy  output 
\citep{TenorioTagleetal2007}. The two solutions are separated in the 
mechanical energy (or cluster mass) {\it vs} size diagram by a critical line, 
the position of which strongly depends on the radiative cooling law applied to
the flow \citep[see][]{TenorioTagleetal2007,TenorioTagleetal2013}. This 
however, as first pointed out by \citet{Silichetal2004}, is not the end of the
story. As one considers more massive clusters of a given size and approaches 
the critical line, the stationary cluster wind becomes strongly radiative, letting its 
temperature drop to $\sim 10^4$ K, closer to the cluster surface.  
Consequently such cluster winds would not present an extended X-ray free 
wind region but instead a recombining and, given the ample supply of UV 
photons, a re-ionized region,  rapidly expanding close to the cluster surface.
    
\begin{figure}[htbp]
\plotone{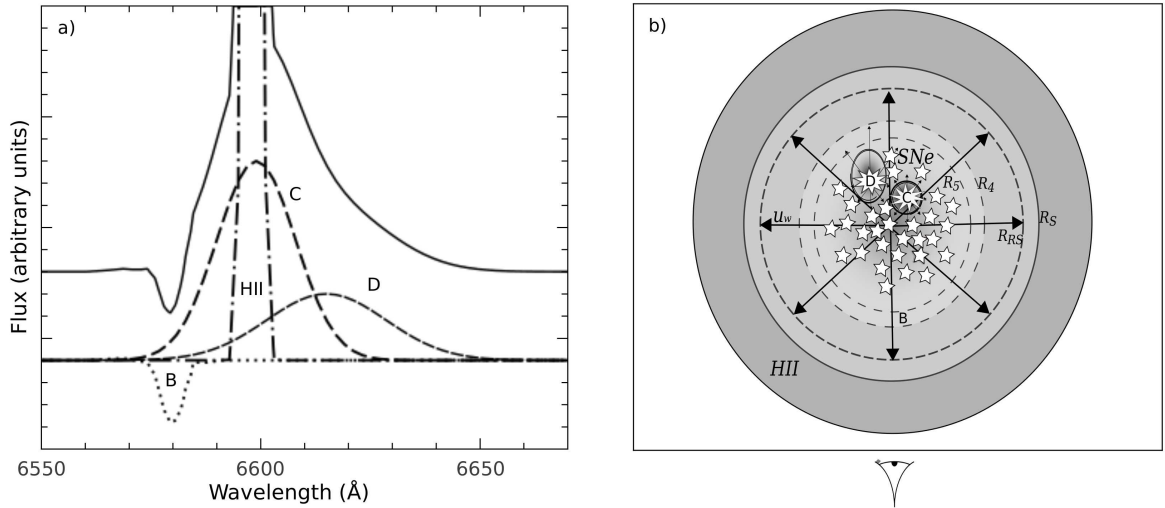}
\caption{The structure of {\it PHL~293B}. Panel a point at the emission and 
absorption line components that lead to the typical ionized H spectra in 
{\it PHL~293B} (solid line). The narrow line, well centered at the galaxy 
redshift, is associated to region HII in panel b. The narrow absorption 
line blue-shifted by some 800~km s$^{-1}$  is here associated to region B and 
the two broad red-shifted components correspond to two SN remnants 
evolving on the far side of the cluster environment. Panel b displays the 
proposed structure of the galaxy and its ionizing cluster (note scales greatly
distorted). Indicated are: the central ionizing cluster that contains two 
rapidly evolving SN remnants (regions C and D in both panels). The high 
velocity cluster wind (large arrows) impacting the galaxy ISM (region HII in 
both panels) and thus supporting a leading shock at a radius $R_S$ and a 
reverse shock at a radius $R_{RS}$. Indicated also is region B  the shell at 
which the stationary cluster wind reaches temperatures 10$^5$ K and 10$^4$ K ($R_5$ and 
$R_4$, respectively) and thus is able to recombine, as well as the observers location.}
\label{fig:1}
\end{figure}

Figure \ref{fig:1} displays the emission and absorption components required to match the
HII spectra of {\it PHL~293B} (see Figure 2 of \citealt[]{Terleetal2014} or 
Figure 3 of \citealt{Izotovetal2011}), together with a sketch showing the 
structure of the galaxy and of its ionizing cluster, as well as the observers 
location. Various regions in the galaxy sketch and in the composed spectra are
labeled (HII and B-D) to indicate their proposed correspondence. In this way, the
narrow emission line component, centered at the redshift of the galaxy and 
showing its low metal abundance, results from the general ionization of the 
galaxy ISM (region HII). The narrow absorption line components are due to a  
fast ($\sim 800$ km s$^{-1}$) strongly radiative cluster wind (region B) and
the broadest components arise from rapidly evolving remnants cause by the
off-centered explosion of the most recent type II supernovae (regions C and 
D).   

\subsection{The cluster wind}
\label{sec:2.1}

In order to reproduce the absorption and emission line components in the spectra of {\it PHL~293B}, 
we have performed several hydrodynamic calculations and made analytical estimates of the feedback generated
within  the central star cluster. In all cases it was assumed that massive stars follow a generalized Schuster 
stellar density distribution \citep{Palousetal2013}; in particular, with the form 
$\rho_* \propto  [1+(r/R_{c})^2]^{-b}$ with $b = 1.5$, where $r$ is the distance from the center and $R_{c}$ is 
the core radius of the stellar distribution. This special case of the Schuster distribution appears as an asymptotic 
case to King's surface formula \citep{King1962} when projected onto the sky \citep{Ninkovic1998}. 
We assume the stationary presence of dust grains within the matter reinserted by massive stars as due to its continuous 
deposition by core-collapse supernovae and use the hydrodynamic treatment described by \citet{TenorioTagleetal2013}. 
We employed the equilibrium cooling function for an optically thin plasma obtained by \citet{Raymondetal1976} for solar 
metallicity and the contribution to the cooling due to gas-dust grain collisions for which we follow the prescription given 
by \citet{DwekWerner1981}. The latter is the dominant source of cooling for a gas with temperatures $T \ge 10^6$ K 
(about two orders of magnitude larger than the gas cooling for a gas in collisional ionization equilibrium, as originally 
calculated by \citet{OstrikerSilk1973}. 

For the calculations we consider clusters close to the critical line in the 
mechanical luminosity (or cluster mass) {\it vs} size and that lead, as 
mentioned in section 2, to strongly radiative winds and thus to recombination 
of the stationary flow close to the cluster surface. In all cases  we fixed 
the cluster half-mass radius, $R_{HM}$,  and the cluster core radius, 
$R_{c}$, and then calculate the corresponding cut-off radius, $R_{SC}$.  Other input parameters are: the star cluster 
mechanical luminosity, $L_{SC}$ and the adiabatic terminal speed $V_{A \infty}$. The mass deposition rate $ \dot{M}$ 
is then: $\dot M = L_{SC}/ V_{A \infty}^2$.  The mechanical luminosity is assumed to scale with the total mass 
of the star cluster $M_{SC}$ as $L_{SC}= 3 \times 10^{40} (M_{SC}/10^6 M_{\sun})$ erg s$^{-1}$ \citep{Leithereretal1999}. 
Following the cluster mass estimates given in \citet{Terleetal2014} we considered star clusters with typical mechanical 
luminosity $L_{SC} = (1-2) \times 10^{40}$~erg s$^{-1}$. Table \ref{tab:1} summarizes the input parameters for our 
models A, B, C and D. Models A and B have the same radii, and a factor of two different mechanical luminosity. 
Models C and D have larger radii and the same  mechanical luminosity as Model B. Models A-C are applicable once 
supernovae begin to take place within the cluster volume ($\sim$ 3 Myr of evolution) and a continuous presence of dust within the reinserted matter is 
established \citep[see][]{TenorioTagleetal2013}. In models A-C, the dust to
gas mass ratio, $Z_d$, was set to $10^{-3}$, dust grains were assumed to be 
spherical and with a radius of $0.1 \mu$m. We also explored, for the sake of 
comparison, a model (D) with the same input parameters as model C, but without 
dust radiative cooling. 
\begin{table}[htp]
\caption{\label{tab:1} Input Parameters}
\begin{tabular}{c c c c c c c }
\hline\hline
Models  & $R_{HM}$ & $R_{c}$ & $R_{SC}$ &   $L_{SC}$ & $V_{A \infty}$ &  $Z_{d}$ \\
       &   (pc) & (pc)   &  (pc) & (erg s$^{-1}$)    &  (km s$^{-1}$) &  $     $ \\ \hline
A      &   1   &  $ 1$   &   $1.46$ & $1 \times 10^{40}$ & 
           1000 & $ 10^{-3} $ \\
B      &   1   &  $ 1$       & $1.46$ & $2 \times 10^{40}$ & 
           1150 & $ 10^{-3} $ \\
C      &   2   &  $ 2$       & $2.62$ & $2 \times 10^{40}$ &
           1000 & $ 10^{-3} $ \\
D      &   2   &  $ 2$       & $2.62$ & $2 \times 10^{40}$ &
           1000 & $       0 $ \\
\hline\hline
\end{tabular}
\end{table}
\begin{figure}[htbp]
\plotone{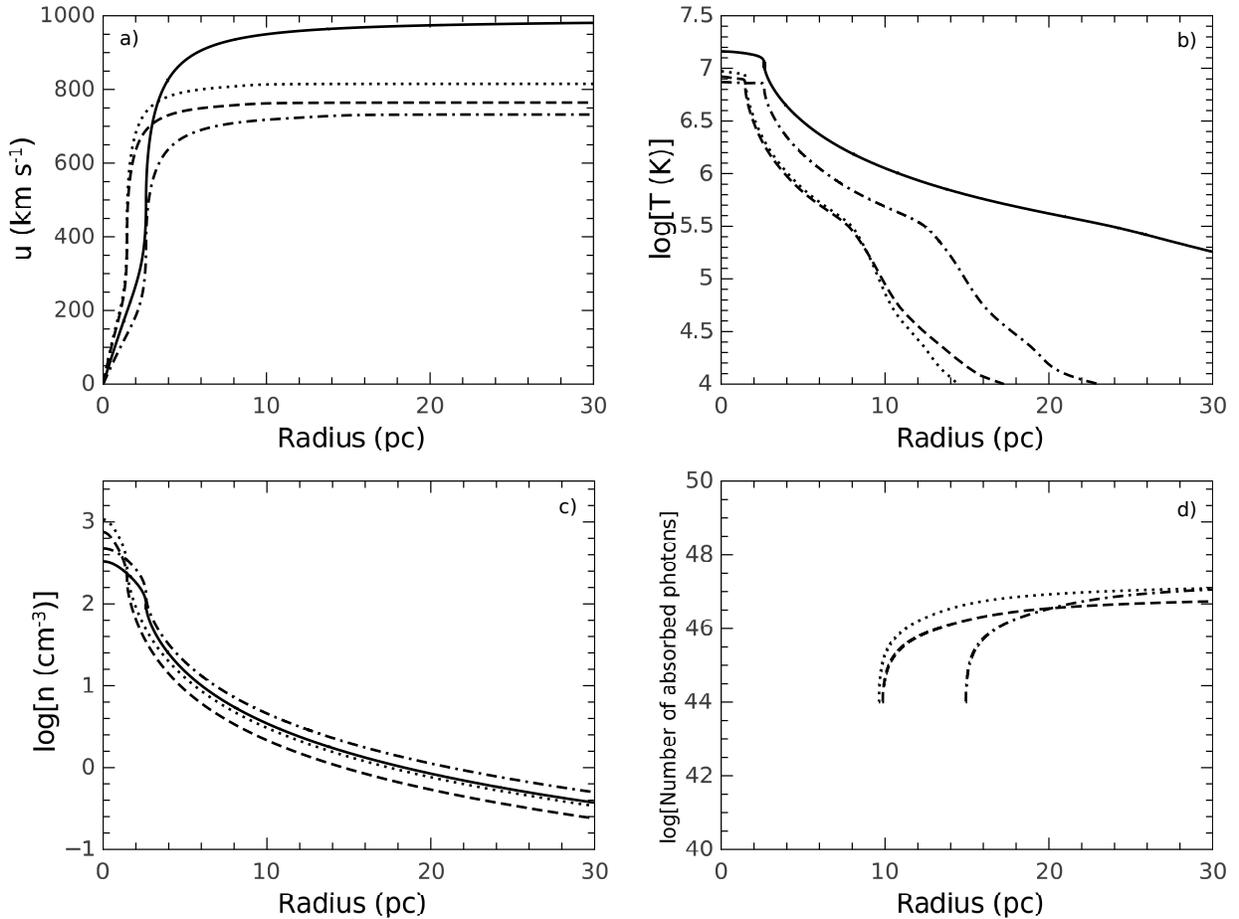}
\caption{The distribution of the hydrodynamical variables and number of 
absorbed photons s$^{-1}$
in the free-wind region. Panels a, b, c and d present the wind velocity, 
temperature, density as a function of distance to the star cluster 
center, respectively, and the number of ionizing photons absorbed per
unit time inside a volume of radius $r$.
The dotted, dashed, dashed-dotted and solid lines display the results of the 
calculations for models A, B, C and D, respectively.}
\label{fig:2}
\end{figure}

Figure \ref{fig:2}, shows the results from the hydrodynamical calculations for models A-D. The dusty models A-C,  
were selected, among many models, because their stationary wind  terminal velocity, $u_{\infty}$, is able to reach a  value  
$\sim 800$ km s$^{-1}$ and further because strong radiative cooling is able to bring the wind temperature down to 
the range $T \sim 10^5 - 10^4$ K, what  allows for H recombination, to a short distance from the star cluster surface. 
The selected models are all for compact star clusters, with a central  wind density value $\sim 10^3$ cm$^{-3}$ and a 
temperature  $\sim 10^7$ K (see Figure \ref{fig:2}). In all cases, once outside the cluster the density drops as $r^{-2}$, 
it reaches values of a few  (cm$^{-3}$)  when the wind 
temperature is $\sim 10^5$ K (at $r$ = $R_5$) between 10 pc and 15 pc, and then values $\sim$ 1 cm$^{-3}$ when the cluster 
wind temperature reaches $10^4$ K, at a distance ($r = R_4$) between $15$ and $22$ pc. It is within this temperature range 
that hydrogen recombination takes place. 

We then used the distributions of density, temperature and velocity presented
in Figure \ref{fig:2} to calculate the outward flux of  ionizing photons 
(cm$^{-2}$~s$^{-1}$) $J = N_{UV}(r) / 4 \pi r^2$ and the degree of ionization 
$x = n_i / (n_i + n_n)$, where $n_i$ and $n_n$ are the ionized and neutral gas
number densities, in the free wind region. The transport of  ionizing 
radiation and the ionization balance equations \citep{Goldsworthy1958} are 
then reduced to:
\begin{eqnarray}
      \label{eq1a}
      & & \hspace{-1.1cm} 
\der{J}{r} = -\frac{2 J}{r} - (1-x) J \frac{\sigma_{a} \rho}{\mu},
      \\[0.2cm]     \label{eq1b}
      & & \hspace{-1.1cm}
\der{x}{r} = (1-x) \frac{\sigma_{a} J}{u} - \frac{\beta \rho}{\mu u} x^2 ,
\end{eqnarray}
where $\sigma_{a} = 6 \times 10^{-18}$~cm$^2$ and $\beta = 3 \times 10^{-10}
T^{-3/4}$~cm$^3$ s$^{-1}$ are the H absorption cross section for ionizing
radiation and the recombination coefficient to  all but the ground level,
respectively, $\mu$ is the mean mass per particle. We solve these equations 
numerically assuming that the ionizing radiation is not depleted in the hot 
wind region until the temperature drops to  $10^5$~K (at $R_5$) and thus that 
the degree of ionization  and 
the total ionizing photon flux are kept $x = 1$ and $N_{UV} = 2.71 \times 
10^{51}$ s$^{-1}$, up to  this radius.

In all our models the degree of ionization $x$ is close  but not equal to
unity. Note also that the number of ionizing photons is large and the wind density has
a moderate value when the wind temperature reaches about $10^5$~K and  
recombination to all energy levels becomes significant. Consequently, in cases
A - C the number of neutral atoms which absorb the ionizing radiation at any 
time $t$ is not negligible. To sustain the value of $x \sim$ 1, some 
10$^{46} - 10^{47}$ ionizing photons s$^{-1}$ are depleted from the radiation 
field to balance recombination within the volume enclosed between $R_5$ and 
$R_4$  (see the last panel of Figure \ref{fig:2}). 
Further out, the density continues to fall in all cases, inhibiting recombination in the outer volumes 
and thus the number of depleted ionizing photons s$^{-1}$ reaches an asymptotic value (see panel d in 
Figure \ref{fig:2}). It is within this region ($R_5 - R_4$) that radiation from the non-ionizing stellar 
continuum, blue-shifted from the emission line centers by some 800 km s$^{-1}$, finds the neutral atoms at 
different energy levels  that produce the weak absorption lines seen in the spectra of  {\it PHL~293B}.    
In the dustless case D the temperature drops to $10^5$~K at a much
larger distance from the star cluster center ($R_5 \approx 37$~pc). By then the wind density has fallen 
sufficiently as to inhibit recombination, therefore this case is not presented on the last panel of Figure \ref{fig:2}.
  
One can also calculate the photoionized ($N(HII)$) and neutral ($N(HI)$)
hydrogen column densities  taking into account that the degree of 
ionization $x$ is close to unity in the whole free wind region: 
$N(HII) = \int_{R_5}^\infty n_5 (R_5/r)^2 dr \approx n_5 R_5$ and  
$N(HI) = \int_{R_5}^\infty n_5 (1 - x) (R_5/r)^2 dr$, 
where $n_5$ is the wind number density at the radius $R_5$. In all models with
dust cooling the calculated values of $N(HII)$ fall in the range $6.6 \times 
10^{19}$~cm$^{-2} < N(HII) < 9.5 \times 10^{19}$~cm$^{-2}$, in good agreement 
with the observational estimate $(1 - 2) \times 10^{20}$~cm$^{-2}$ by
\citet{Terleetal2014}. The value of $(1 - x)$ changes from 0 at $r = R_5$ 
to the asymptotic value of $2.3 \times 10^{-7}$ (model A), $3.2 \times 
10^{-7}$ (model B) and $4.7 \times 10^{-7}$ (model C) at a distance $R_4$
from the star cluster center. The upper limit to the neutral 
hydrogen column density then is: $N(HI) = (1 - x)_a \int_{R_5}^\infty 
n_5 (R_5/r)^2 dr = (1 - x)_a N(HII)$, where $(1 - x)_a$ is the asymptotic 
value of the $(1 - x)$ factor. This leads to  neutral hydrogen column densities
within the range $1.5 \times 10^{13}$~cm$^{-2} < N(HI) < 4.5 \times 
10^{13}$~cm$^{-2}$ which is also in good agreement with the 
observational estimates \citep[see section 1 and  ][]{Terleetal2014}.
One can compare now these results
with the dustless model D, which exhibits a quasi-adiabatic behavior, to note 
that despite the similar density and velocity distributions, model D does not 
present the drastic temperature fall at a short distance from the star cluster
center and therefore it would not lead to the high velocity blue-shifted 
absorption features
 predicted for models A-C. 

Finally, we have used the calculated model profiles to estimate the expected 
diffuse X-ray emission:
\begin{equation}
      \label{eq2}
L_X = 4 \pi \int_0^{R_{out}} r^2 n_e n_i \Lambda_X(T,Z) {\rm d}r ,
\end{equation}
where $n_e(r)$ and $n_i(r)$ are the electron and ion number densities, 
$\Lambda_X(T,Z)$ is the X-ray emissivity used by Strickland \& Stevens 
(2000), $Z$ is the metallicity of the X-ray plasma in solar units 
and $R_{out}$ marks the X-ray cut-off radius (the radius where the 
temperature in the wind drops below $T_{cut} \approx 5 \times 10^5$ K). 
For $Z=1$ we found that in cases A and D the 0.3 keV - 2.0 keV 
diffuse luminosity ($L_{X} = 4.5 \times 10^{38}$~erg s$^{-1}$ and
$ 4.0 \times 10^{38}$~erg s$^{-1}$, respectively) are in  reasonable
agreement with the observed upper limit of $\sim 2.2 \times 10^{38}$~erg 
s$^{-1}$ \citep{Terleetal2014} whereas models B and C predict a somewhat 
larger X-ray emission: $L_{X} = 8.8 \times 10^{38}$~erg 
s$^{-1}$ and $ 1.2 \times 10^{39}$~erg s$^{-1}$, respectively.
$L_X$ scales almost linearly with $Z$ and brings it into better agreement
with the observational limits if the metallicity of the hot gas is subsolar.
For example, in model A it falls to $L_X \approx 1.5 \times 10^{38}$~erg 
s$^{-1}$ when $Z = 0.3$.

A hot dusty plasma radiates mainly in the X-ray and IR regimes. Thus one can 
obtain the upper limit for the IR luminosity expected in cases A, B and C
if one compares the star cluster mechanical luminosity $L_{SC}$ with the energy 
flux through a sphere of radius $r = R_{out}$ at which the X-ray emission 
vanishes and assume that all energy lost within this volume is radiated
away either in the IR or in the X-ray regime: 
\begin{equation}
      \label{eq3}
L_{IR} \approx L_{SC} - L_X - 4 \pi \rho u R^2_{out} 
       \left(\frac{u^2}{2} + \frac{\gamma}{\gamma-1}\frac{p}{\rho}\right) ,
\end{equation}
where $\gamma = 5/3$ is the ratio of specific heats and all hydrodynamical
variables (the density $\rho$, velocity $u$ and the value of thermal pressure
$p$) are calculated at $r = R_{out}$. This leads to IR luminosities
$L_{IR} = 3.6 \times 10^{39}$~erg s$^{-1}$, $8.9 \times 10^{39}$~erg 
s$^{-1}$ and $7.9 \times 10^{39}$~erg s$^{-1}$   
in cases A, B and C, respectively.
The dusty wind models thus lead to  IR luminosities which exceed the
cluster wind diffuse X-ray emission by about an order of magnitude ($L_{IR}/L_X =$
8.0, 10.1 and 6.8 in models A, B and C, respectively). In the dustless
case D the hot wind cools only through  X-ray radiation. $L_X$ presents $\sim 87$\% 
of the total radiated energy in this case.

\subsection{The broad emission lines}
\label{sec:2.2}

The broad emission components in the spectra of {\it PHL~293B}, depicted for example  in Figure 2 of \citet{Terleetal2014}, 
are here ascribed to  two rapidly evolving SN remnants resultant from the most recent type II supernova explosions that had 
taken place within the ionizing cluster volume (see Figure \ref{fig:1}). Furthermore, both explosions have occurred on the 
far side of the cluster, to account for the intrinsic redshifted emission of the evolving remnants that should expand with 
slower velocities  into regions of higher density (towards us and the cluster center) and faster into lower density regions 
(towards the cluster edge). As we know the FWZI of the broadest component ($\sim 4000$ km s$^{-1}$) and 
have a good estimate of the background density (see Figure \ref{fig:2}, panel c) one can
estimate the size of such remnants as well as their cooling time which mark 
the time when photoionization will lead to their strong emission at optical 
wavelengths. The size $R_S$ of such SN-driven shells from the energy 
conservation relation is:
\begin{equation}
\label{eq5}
R_S^3 =  \frac{3 \alpha E_{SN}}{2 \pi \rho V^2_{exp}}
\end{equation}
where $E_{SN} = 10^{51}$ erg is the energy of the explosion, $\alpha$ is the 
fraction of the explosion energy which is transformed into  kinetic energy of 
the swept-up gas ($\alpha \approx 0.3$ during the Sedov phase), and $M = 4 
\pi \rho R^3_S / 3$, $V_{exp}$ and $R_S$ are the mass, expansion velocity and 
radius of the swept-up gas, respectively.
Taking $V_{exp}$ = 1/2 FWZI =  2$\times 10^3$ km s$^{-1}$ and  the average gas
density within the star cluster volume  as $\rho \sim 10^{-21}$ g cm$^{-3}$ 
(see Figure \ref{fig:2}) results into a remnant size $R_S \sim 0.5$~pc, which warrants a
strong SN remnant evolution within the cluster volume.

From the jump conditions across a strong shock one can derive the temperature 
of the swept up gas: $T_S = 1.3 \times 10^7 (V_{exp}/10^3$ km s$^{-1})^2$, 
required to estimate the cooling rate, and the shocked gas density, $n_S$ 
(cm$^{-3}$), equal to 4 times the background density. The above relations 
yield a shocked gas temperature $T_S = 5.2 \times 10^7$ K and a shocked gas 
density $n_S$ $\sim 2 \times 10^3$ cm$^{-3}$.  These together with the value 
of the cooling rate, $\Lambda$ $\sim 10^{-20} $ erg cm$^3$ s$^{-1}$ 
\citep[see Figure 2 of][]{TenorioTagleetal2013}  lead to the cooling time:
\begin{equation}
t_{\Lambda} = \frac{3 k T_S}{n_S \Lambda}\approx \mbox{a few } 100 \mbox{ yr}
\end{equation}
Thus the supernova remnants expanding within the cluster environment 
experience a strong evolution and after cooling become exposed to the cluster 
ionizing radiation which makes them shine at optical wavelengths. 
The number of ionizing photons absorbed by such a remnant is
\begin{equation}
N_{abs} = 4 \pi F R^2_S = 4 \pi q_{UV} R_{SC} R^2_S = 3 X N_{UV} 
          (R_S/R_{SC})^2 ,
\end{equation}
where $F$ is the flux of ionizing photons per unit area of the supernovae 
remnant, $q_{UV} = 3 N_{UV} / 4 \pi R^3_{SC}$ and the factor $X$
is $1/4 < X < 1/3$ \citep[see][]{Palousetal2014}.
If $X=1/3$, the ratio of broad to narrow emission lines components is:
\begin{equation}
L_{broad}/L_{narrow} = (R_S/R_{SC})^2/[1 - (R_S/R_{SC})^2] . 
\end{equation}
This ratio varies with the supernova remnant size. For $R_S = 0.5$~pc and
$R_{SC} = 1.46$~pc it is about 13\%. Given the fact that the remnant is 
expected to be elongated, as it expands into gas with different densities 
inside the cluster volume, it may easily reach the 20\% value measured by
\citet{Terleetal2014}.

\subsection{The narrow emission line components}
\label{sec:2.3}

As depicted in Figure \ref{fig:1}, the narrow emission line components in the spectra of {\it PHL~293B} arise from 
the extended HII region that surrounds the central cluster. At first glance, the narrowness of the emission lines and 
the low metallicity derived from the various line intensity ratios,  
imply that such an extended HII region is still unaffected by the mechanical energy of the central cluster. Note 
however that the interaction of the cluster wind with the galaxy background would unavoidably lead to the 
formation of a hot ($\sim 8 \times 10^6$ K) wind driven bubble and its surrounding supershell.
The hot wind driven bubble results from the thermalization of the wind at a reverse shock and the surrounding super 
shell
from the matter swept up by the leading shock. One expects the velocity of the supershell to decay as the evolution 
proceeds and more mass is gathered while it acquires a larger dimension. Estimates of both the expansion speed and 
the size of a supershell at a given time, can be made if one knows the cluster mechanical luminosity ($L_{SC}$) the 
age ($t_{SC}$) and the background density ($n_0$). And thus, if one uses the data for {\it PHL~293B} i.e. 
an $L_{SC} \sim 10^{40}$ erg s$^{-1}$, an evolution time $t_{SC} \sim 5$ Myr and a background density 
$n_0 \sim 100$ cm$^{-3}$ \citep{Terleetal2014}, equations 1 and 2 of \citet{MacLowMcCray1988}, for example, 
lead to an expansion speed of about 20 km s$^{-1}$ and a radius of 176 pc. This implies a supershell well contained 
within the galaxy. However, such an expansion speed, similar to the sound speed in the photoionized region 
($T_{HII} \sim  15600$ K, see \citealt{Izotovetal2011}) also implies that the shell is presently being disrupted by 
the random speed of motions in the ISM, what makes the supershell spectroscopically undetectable in {\it PHL~293B}.

\section{Conclusions}
\label{sec:3}

A thorough study of different observational data sets of {\it PHL~293B} that 
span over the last two decades \citep{Terleetal2014} has shown a lack of 
variability, what rules out the possibility of a luminous blue variable star
or a type IIn SN as the primary source of the spectra detected in this galaxy.
The idea of an  expanding ($V_{exp} \sim $ 800 km s$^{-1}$) supershell causing
the absorption components can also be discarded  as its dynamical time will 
place it way out of the dwarf galaxy.  On the other hand, we have shown here
that a powerful, strongly radiative, stationary cluster wind is able to 
produce a stationary high velocity recombining layer that causes blue-shifted
absorption features similar to those observed in the spectra of 
{\it PHL~293B}.  We have taken into consideration the mass, 
mechanical luminosity, age and flux of ionizing photons from the central 
cluster of {\it PHL~293B} to built a grid of possible models of the stationary
cluster wind. From these we have  selected  as favorites some of those that 
lead to a terminal velocity around 800 km s$^{-1}$  and that cause a rapidly 
recombining stationary layer close to the cluster surface. The dusty cluster 
wind models, with strong radiative cooling, predict an ionized and neutral gas
column density as well as a diffuse X-ray emission in very good agreement with
the observed values. We thus claim that the absorption features {\it PHL~293B}
present the first observational evidence for a strongly radiative stationary 
cluster wind, as originally discussed in \citet{Silichetal2004}.

We have also shown that the impact of the cluster wind on the galactic 
environment leads to an expanding supershell
which after 5 Myr of evolution presents a velocity similar to the sound speed of the photoionized gas and has
become spectroscopically undetectable.

The remnants of two recent type II supernova explosions, that burst on the far side of the stellar cluster environment, 
have been claimed in our model as responsible  for the broad redshifted emission components in the H spectra of 
{\it PHL~293B}. We have shown that even the fastest of these remnants undergoes a rapid evolution, reaching its cooling 
time while evolving well within the cluster volume. After such a time, the remnants become a target of the stellar 
ionizing radiation that allows us to trace them as the fastest structures in the galaxy.     

\section{Acknowledgments}

We thank our anonymous referee for many suggestions that greatly improved the
clarity of the paper. This study has been supported by CONACYT - M\'exico, 
research grants 167169, 131913 and 2005-4984, 2008-103365 and by the Spanish 
Ministry of Science and Innovation under the collaboration ESTALLIDOS (grants 
AYA2007-67965-C03-01, AYA2007-67965-C03-0, AYA2010-21887-C04-03 and 
AYA2010-21887-C04-04). GTT also acknowledges the C\'atedra Jes\'us Serra at 
the Instituto de Astrof\'isica de Canarias (IAC, Tenerife Spain) and SS
appreciates the hospitality of his colleagues during his stay at the IAC.

\bibliographystyle{apj}
\bibliography{PHL293B}

\begin{thebibliography}{}
\expandafter\ifx\csname natexlab\endcsname\relax\def\natexlab#1{#1}\fi

\bibitem[{{Asplund} {et~al.}(2009){Asplund}, {Grevesse}, {Sauval}, \&
  {Scott}}]{Asplundetal2009}
{Asplund}, M., {Grevesse}, N., {Sauval}, A.~J., \& {Scott}, P. 2009, \araa, 47,
  481

\bibitem[{{Cair{\'o}s} {et~al.}(2001){Cair{\'o}s}, {V{\'{\i}}lchez},
  {Gonz{\'a}lez P{\'e}rez}, {Iglesias-P{\'a}ramo}, \& {Caon}}]{Cairosetal2001}
{Cair{\'o}s}, L.~M., {V{\'{\i}}lchez}, J.~M., {Gonz{\'a}lez P{\'e}rez}, J.~N.,
  {Iglesias-P{\'a}ramo}, J., \& {Caon}, N. 2001, \apjs, 133, 321

\bibitem[{{Dwek} \& {Werner}(1981)}]{DwekWerner1981}
{Dwek}, E., \& {Werner}, M.~W. 1981, \apj, 248, 138

\bibitem[{{French}(1980)}]{French1980}
{French}, H.~B. 1980, \apj, 240, 41

\bibitem[{{Geha} {et~al.}(2006){Geha}, {Blanton}, {Masjedi}, \&
  {West}}]{Gehaetal2006}
{Geha}, M., {Blanton}, M.~R., {Masjedi}, M., \& {West}, A.~A. 2006, \apj, 653,
  240

\bibitem[{{Goldsworthy}(1958)}]{Goldsworthy1958}
{Goldsworthy}, F.~A. 1958, Reviews of Modern Physics, 30, 1062

\bibitem[{{Izotov} {et~al.}(2011){Izotov}, {Guseva}, {Fricke}, \&
  {Henkel}}]{Izotovetal2011}
{Izotov}, Y.~I., {Guseva}, N.~G., {Fricke}, K.~J., \& {Henkel}, C. 2011, \aap,
  533, A25

\bibitem[{{Izotov} {et~al.}(2007){Izotov}, {Thuan}, \&
  {Stasi{\'n}ska}}]{Izotovetal2007}
{Izotov}, Y.~I., {Thuan}, T.~X., \& {Stasi{\'n}ska}, G. 2007, \apj, 662, 15

\bibitem[{{King}(1962)}]{King1962}
{King}, I. 1962, \aj, 67, 471

\bibitem[{{Kunth} \& {{\"O}stlin}(2000)}]{KunthOstlin2000}
{Kunth}, D., \& {{\"O}stlin}, G. 2000, \aapr, 10, 1

\bibitem[{{Leitherer} {et~al.}(1999){Leitherer}, {Schaerer}, {Goldader},
  {Gonz{\'a}lez Delgado}, {Robert}, {Kune}, {de Mello}, {Devost}, \&
  {Heckman}}]{Leithereretal1999}
{Leitherer}, C., {Schaerer}, D., {Goldader}, J.~D., {et~al.} 1999, \apjs, 123,
  3

\bibitem[{{Mac Low} \& {McCray}(1988)}]{MacLowMcCray1988}
{Mac Low}, M.-M., \& {McCray}, R. 1988, \apj, 324, 776

\bibitem[{{Ninkovic}(1998)}]{Ninkovic1998}
{Ninkovic}, S. 1998, Serbian Astronomical Journal, 158, 15

\bibitem[{{Ostriker} \& {Silk}(1973)}]{OstrikerSilk1973}
{Ostriker}, J., \& {Silk}, J. 1973, \apjl, 184, L113

\bibitem[{{Palou{\v s}} {et~al.}(2014){Palou{\v s}}, {W{\"u}nsch}, \&
  {Tenorio-Tagle}}]{Palousetal2014}
{Palou{\v s}}, J., {W{\"u}nsch}, R., \& {Tenorio-Tagle}, G. 2014, \apj, 792,
  105

\bibitem[{{Palou\v{s}} {et~al.}(2013){Palou\v{s}}, {W\"unsch},
  {Mart\'inez-Gonz\'alez}, {Hueyotl-Zahuantitla}, {Silich}, \&
  {Tenorio-Tagle}}]{Palousetal2013}
{Palou\v{s}}, J., {W\"unsch}, R., {Mart\'inez-Gonz\'alez}, S., {et~al.} 2013,
  \apj, 772, 128

\bibitem[{{Raymond} {et~al.}(1976){Raymond}, {Cox}, \&
  {Smith}}]{Raymondetal1976}
{Raymond}, J.~C., {Cox}, D.~P., \& {Smith}, B.~W. 1976, \apj, 204, 290

\bibitem[{{Silich} {et~al.}(2004){Silich}, {Tenorio-Tagle}, \&
  {Rodr{\'{\i}}guez-Gonz{\'a}lez}}]{Silichetal2004}
{Silich}, S., {Tenorio-Tagle}, G., \& {Rodr{\'{\i}}guez-Gonz{\'a}lez}, A. 2004,
  \apj, 610, 226

\bibitem[{{Tenorio-Tagle} {et~al.}(2013){Tenorio-Tagle}, {Silich},
  {Mart{\'{\i}}nez-Gonz{\'a}lez}, {Mu{\~n}oz-Tu{\~n}{\'o}n}, {Palou{\v s}}, \&
  {W{\"u}nsch}}]{TenorioTagleetal2013}
{Tenorio-Tagle}, G., {Silich}, S., {Mart{\'{\i}}nez-Gonz{\'a}lez}, S., {et~al.}
  2013, \apj, 778, 159

\bibitem[{{Tenorio-Tagle} {et~al.}(2007){Tenorio-Tagle}, {W{\"u}nsch},
  {Silich}, \& {Palou{\v s}}}]{TenorioTagleetal2007}
{Tenorio-Tagle}, G., {W{\"u}nsch}, R., {Silich}, S., \& {Palou{\v s}}, J. 2007,
  \apj, 658, 1196

\bibitem[{{Terlevich} {et~al.}(2014){Terlevich}, {Terlevich}, {Bosch},
  {D\'iaz}, {Hagele}, {Cardaci}, \& {Firpo}}]{Terleetal2014}
{Terlevich}, R., {Terlevich}, E., {Bosch}, G., {et~al.} 2014, \mnras, in press

\end{thebibliography}

\end{document}